\documentclass[aps,preprint,a4paper,showpacs,showkeys,superscriptaddress]{revtex4-1}
\ifx\pdfoutput\undefined
\usepackage[usenames]{color} 
\usepackage[dvips]{graphicx}
\else
\usepackage{color} 
\definecolor{BrickRed}{rgb}{0.85,0.15,0.25}
\definecolor{MidnightBlue}{rgb}{0,0.45,0.85}
\definecolor{ForestGreen}{rgb}{0,0.85,0.45}
\usepackage{physics}
\usepackage{graphicx}
\usepackage{epstopdf}
\fi
\usepackage{subfigure}
\usepackage{latexsym,amsmath,amssymb}
\usepackage{hyperref}     
\hypersetup{colorlinks,%
  citecolor=blue,%
  linkcolor=cyan,%
  pdftex}
\allowdisplaybreaks

\usepackage{acronym}

\usepackage{umoline}
\newsavebox\CBox

\begin{document}


\title{Investigation of the black-hole quantum atmosphere from the effective proper temperature}

\author{Sojeong Cheong}%
\email[]{jsquare@sogang.ac.kr}%
\affiliation{Center for Quantum Spacetime, Sogang University, Seoul 04107, Republic of Korea}%

\author{Wontae Kim}%
\email[]{wtkim@sogang.ac.kr}%
\affiliation{Center for Quantum Spacetime, Sogang University, Seoul 04107, Republic of Korea}%
\affiliation{Department of Physics, Sogang University,
Seoul 04107, Republic of Korea}%

\date{\today}

\begin{abstract}

Hawking radiation may be regarded as originating at the event horizon; however, its
spatial origin can instead be distributed over a finite region outside the horizon.
In this paper, using an effective proper temperature, we investigate the quantum atmosphere within a tractable model based on the two-dimensional radial sector of $D$-dimensional Schwarzschild black holes.
In the Hartle--Hawking state, the proper temperature is derived from the first law of thermodynamics in the presence of the conformal anomaly associated with
Hawking radiation. In the Unruh state, we decompose the proper temperature
into two chiral temperatures, $T_{\rm L}$ and $T_{\rm R}$, associated with the ingoing and outgoing fluxes, respectively.
Demonstrating $T_{\rm R}$ as the effective proper temperature characterizing the outgoing Hawking flux with the proper temperature in the Hartle-Hawking state, we define the atmospheric radius effectively as the radial position at which the effective proper temperature attains its maximum.
We then numerically compute the atmospheric radius of the quantum atmosphere for various spacetime dimensions and show that it
decreases monotonically as the number of spacetime dimensions increases.
In the large-dimensional limit, we find that the atmospheric radius remains separated from the horizon by a finite radial factor, indicating that the quantum atmosphere can persist as an extended exterior region in any dimension.
\end{abstract}


\keywords{Hawking radiation, quantum atmosphere, proper temperature, conformal anomaly, Stefan--Boltzmann law}

\maketitle


\acrodef{AMPS}[AMPS]{Almheiri, Marolf, Polchinski and Sully}
\acrodef{CGHS}[CGHS]{Callan-Giddings-Harvey-Strominger}
\acrodef{RST}[RST]{Russo-Susskind-Thorlacius}

\section{Introduction}
\label{sec:intro}

Hawking radiation~\cite{Hawking:1974sw} may play a crucial role as a possible carrier of black-hole information.
In connection with the black-hole information paradox, black-hole complementarity, proposed as a possible resolution to the paradox,
depends on the observer and the black-hole vacuum state under consideration; however, no observer should encounter physical observations contradictory to those of another observer~\cite{Susskind:1993if}.
Subsequently, the firewall paradox emphasized this issue by arguing that
the validity of semiclassical quantum field theory outside the horizon
and the equivalence principle cannot be simultaneously maintained~\cite{Almheiri:2012rt}.
One of the proposals for resolving this conflict is that
an infalling observer crossing the horizon after the Page time~\cite{Page:1993wv}
would encounter a firewall composed of high-frequency quanta with trans-Planckian energies. This proposal has led to extensive discussions and various alternatives to the firewall scenario~\cite{Braunstein:2009my,Bousso:2012as, Nomura:2012sw, Susskind:2012rm, Hossenfelder:2012mr, Giddings:2013kcj,
Almheiri:2013hfa, Hutchinson:2013kka, Freivogel:2014dca, Israel:2015ava}.

One of the essential questions behind these discussions is where the outgoing Hawking quanta are effectively produced.
At first glance, the Hawking temperature measured at infinity is gravitationally blueshifted for a local observer at a finite distance~\cite{Tolman:1930zza},
and so Hawking flux seems to be generated only at, or infinitesimally close to, the horizon.
However, such a naive near-horizon picture is in contrast with the regularity condition
for the outgoing flux of the evaporating black hole in the Unruh state~\cite{Unruh:1976db}.
As a matter of fact,
the renormalized stress tensor in the Unruh state suggests that the outgoing Hawking radiation cannot be attributed solely to Planckian excitations arbitrarily close to the horizon, but instead becomes significant over an extended exterior region~\cite{Unruh:1994zw}.

The idea of a black-hole quantum atmosphere was elaborated by Giddings, who proposed that Hawking radiation is not regarded as being produced in an arbitrarily thin region very near the horizon,
but instead as emerging from an extended region outside the horizon~\cite{Giddings:2015uzr}.
In this picture, the transition of ingoing modes into outgoing Hawking flux can occur over the quantum atmosphere, which is compatible with nonviolent scenarios for the information loss paradox without introducing firewall-like high-energy excitations at the horizon~\cite{Giddings:2006sj}.
This interpretation was supported by studying the radiating flux and the stress tensor~\cite{Kim:2016iyf,Dey:2017yez,Dey:2019ugf}, where the Hawking quanta were argued to originate from the quantum atmosphere around the black hole.
More recently, related studies have also explored various aspects of the quantum atmosphere and near-horizon physics~\cite{Eune:2019aat,Shallue:2025zto,Zhang:2025xes,Liu:2026exs,Ming:2026lud}.

Alternatively, for an asymptotic observer, an effective atmospheric radius can be investigated by comparing the total power of Hawking radiation measured at infinity with the Stefan--Boltzmann power of a blackbody emitter at the Hawking temperature~\cite{Giddings:2015uzr}.
From numerical calculations of the power of Hawking radiation~\cite{Page:1976df,Elster:1983pk,Cardoso:2005vb},
the atmospheric radii in four and higher dimensions were computed
by using the $D$-dimensional Stefan--Boltzmann law~\cite{Hod:2016hdd}:
the atmospheric radius of $D$-dimensional Schwarzschild black holes decreases with the number of spacetime dimensions and eventually collapses toward the horizon in the large-dimensional limit.
It raises the question: is it possible to quantify the characteristic feature of the quantum atmosphere locally?
If such a local characterization is possible, the dimensional dependence of the quantum atmosphere can be clarified further, including whether
its notion can persist in arbitrary dimensions or not.

One possible way to characterize the quantum atmosphere locally is
to construct an effective proper temperature for the outgoing flux from black holes and identify the position of its maximum with the atmospheric radius.
In two and four dimensions, such an effective proper temperature was obtained explicitly from the renormalized stress tensor~\cite{Gim:2015era,Eune:2015xvx},
where the maximum of the effective proper temperature in four dimensions was shown to be shifted closer to the horizon than that in two dimensions.
To extend this analysis to arbitrary higher dimensions, one would need sufficient information about the renormalized stress
in order to define the corresponding $D$-dimensional effective proper temperature.
In higher dimensions, however, the full renormalized stress tensor is unfortunately unknown in general.
Therefore, as a tractable model for investigating the higher-dimensional dependence,
we consider the two-dimensional radial sector of the $D$-dimensional Schwarzschild geometry.
In this approach, the renormalized stress tensor can be treated explicitly, allowing us to explore how the effective proper temperature and the corresponding atmospheric radius depend on the spacetime dimension.

The key ingredient in our calculations is to decompose the proper temperature
into two chiral temperatures: $T_\text{L}$ and $T_\text{R}$, corresponding to the ingoing and outgoing fluxes in the Unruh state, respectively.
Identifying $T_\text{R}$ with an effective proper temperature $T_{\rm eff}$ characterizing the outgoing Hawking flux for the quantum atmosphere,
we define the atmospheric radius as the radial position at which the effective proper temperature attains its maximum. We then numerically compute the atmospheric radius of the quantum atmosphere for various spacetime dimensions and show that it
decreases monotonically as the number of spacetime dimensions increases.
In the large-dimensional limit, the atmospheric
radius does not collapse onto the horizon but remains separated from the horizon by a finite radial factor.
Consequently, it implies that the spatial origin of Hawking radiation can persist as an extended exterior region even when the number of spacetime dimensions becomes large.

The organization of this paper is as follows.
In Sec.~\ref{sec:SBlaw},
we introduce the semiclassical stress tensor for $N$ massless scalar fields
on a general two-dimensional static black-hole background.
In thermal equilibrium, we derive the modified Stefan--Boltzmann law in the presence of the conformal anomaly related to the Hawking radiation and then obtain the corresponding proper temperature.
In Sec.~\ref{sec:temp},
the proper temperature in the Unruh state can be decomposed into $T_{\rm L}$ and $T_{\rm R}$.
We prove that the effective proper temperature for the outgoing flux is the same as the proper temperature in the Hartle--Hawking state.
In Sec.~\ref{sec:atmosphere}, we apply the effective proper temperature to the two-dimensional radial sector of $D$-dimensional Schwarzschild black holes
in order to investigate the dimensional dependence of the quantum atmosphere.
Finally, we present our conclusion and discussion in Sec.~\ref{sec:conclusion}.

\section{Modified Stefan--Boltzmann law}
\label{sec:SBlaw}

For a general two-dimensional static geometry, the line element is given by
\begin{equation} \label{eq:2d_metric}
    \dd s^2= - f(r) \dd t^2 + \frac{1}{f(r)}\dd r^2,
\end{equation}
which can be written in the
conformal gauge: $\dd s^2= -e^{2\rho{(\sigma^+, \sigma^-)}} \dd \sigma^+ \dd \sigma^- = -f(r) \dd \sigma^+ \dd \sigma^- $.
For a local observer, we define $u^\mu$ as the two-velocity and $n^\mu$ as a spacelike unit vector, satisfying $u^\mu u_\mu=-1$, $n^\mu n_\mu=1$, and $n^\mu u_\mu=0$, explicitly written as $u^\pm = \frac{1}{\sqrt{f(r)}}$ and $n^\pm =\pm \frac{1}{\sqrt{f(r)}}$.
Then, one can obtain the local proper quantities such as energy density, pressure, and flux:
\begin{equation} \label{eq:epf}
    \varepsilon=\langle T_{\mu\nu}\rangle u^\mu u^\nu ,\quad p=\langle T_{\mu \nu} \rangle n^\mu n^\nu,\quad {\cal{F}}=-\langle T_{\mu\nu} \rangle u^\mu n^\nu
\end{equation}
in the observer’s orthonormal frame,
where $\langle T_{\mu\nu}\rangle$ is the vacuum expectation value of the stress tensor for the $N$ massless scalar fields on the background \eqref{eq:2d_metric}.

Assuming local thermal equilibrium, the first law of thermodynamics for the local observer can be written as
$\dd U =T\dd S-p\dd V$, where $U$, $T$, $S$, and $V$ are internal energy,
temperature, entropy, and volume of a system in thermal equilibrium, respectively.
From the Maxwell relation $(\frac{\partial S}{\partial V} )_T = (\frac{\partial p}{\partial T})_V$,
the thermodynamic first law
can be expressed by
\begin{equation} \label{eq:thermo0}
    \varepsilon = T \left(\frac{\partial p}{\partial T}\right)_V - p,
\end{equation}
where $U=\int \varepsilon \dd V$ and $T$ is the local proper temperature
of the system at $r$.
In general, the trace of the stress tensor is quantum-mechanically non-vanishing due to the conformal anomaly, which is
formally written as $\langle T^\mu_\mu \rangle
 =-\varepsilon +p$, and thus, Eq.~\eqref{eq:thermo0} can be written as
\begin{equation} \label{eq:thermo1}
 T \left(\frac{\partial \varepsilon}{\partial T}\right)_V -2\varepsilon =   \langle T^\mu_\mu \rangle,
\end{equation}
where we used the fact that the conformal anomaly is temperature-independent~\cite{BoschiFilho:1991xz}.
If the stress tensor were traceless~\cite{Tolman:1930zza}, the familiar Stefan--Boltzmann law $\varepsilon=p=\gamma T^2$ would be
obtained as a solution of the thermodynamic first law~\eqref{eq:thermo1};
however, in the presence of the conformal anomaly responsible for the Hawking radiation~\cite{Christensen:1977jc},
the Stefan--Boltzmann law should be modified.

In the one-loop approximation, varying the Polyakov effective action of the $N$ massless scalar fields with respect
to the metric in the conformal gauge yields~\cite{Polyakov:1987zb}
\begin{align}
\langle T_{\pm\pm}\rangle &=  \hat{T}_{\pm\pm}+ t_\pm(\sigma^\pm), \label{eq:T++beta}\\
\langle T_{+-}\rangle &= \hat{T}_{+-},\label{eq:T+-beta}
\end{align}
where $ \hat{T}_{\pm\pm} = -\frac{N}{12\pi} \left[ (\partial_\pm \rho)^2 - \partial^2_\pm \rho \right]$, $ \hat{T}_{+-} = -\frac{N}{12\pi} \partial_+ \partial_- \rho$, and $t_\pm(\sigma^\pm)$ are arbitrary integration functions from the non-locality of the Polyakov effective action.
For the thermal state, the integration functions on the static background will reduce to constants as $ t_\pm(\sigma^\pm)
= t_\pm(\beta)$, whose values are
determined by the specific choice of black-hole state such as the Hartle--Hawking state~\cite{Hartle:1976tp,Israel:1976ur}.
The thermal information is carried only by the terms of $t_\pm$ in Eq.~\eqref{eq:T++beta}, whereas $ \hat{T}_{\pm\pm}$ in Eq.~\eqref{eq:T++beta} and $ \hat{T}_{+-}$ in Eq.~\eqref{eq:T+-beta} are determined geometrically through the renormalization, independently of the temperature of the heat bath.
In Eq.~\eqref{eq:epf}, the energy density, pressure, and flux for the local observer are expressed as
\begin{align} \label{eq:epf2}
    \varepsilon = \hat{\varepsilon} + t(r) , \quad
    p = \hat{p} + t(r), \quad
    {\cal{F}}  =
    -\frac{1}{f(r)} \left( t_+(\beta) - t_-(\beta) \right),
\end{align}
where $\hat{\varepsilon} = \frac{1}{f} \left(  \hat{T}_{++} +  \hat{T}_{--} +  2 \hat{T}_{+-} \right)$,
$\hat{p} = \frac{1}{f} \left(\hat{T}_{++} + \hat{T}_{--} - 2 \hat{T}_{+-}  \right)$,
and the temperature-dependent part of the proper quantities for the local observer is given by
\begin{equation} \label{eq:t_def}
    t(r) = \frac{1}{f(r)} \left(  t_+(\beta) + t_-(\beta) \right).
\end{equation}
In addition, the conformal anomaly can be written as
\begin{equation} \label{eq:trace_rel}
    \langle T^\mu_\mu \rangle = -\hat{\varepsilon} + \hat{p}.
\end{equation}
It is worth noting that the conformal anomaly is a local geometric quantity, which is independent of the thermal contribution~\cite{BoschiFilho:1991xz}.

Now, in order to obtain the proper temperature in thermal equilibrium,
we require the proper energy density and pressure in Eq.~\eqref{eq:epf2}
to be compatible with the thermodynamic first law~\eqref{eq:thermo1}.
Solving Eq.~\eqref{eq:thermo1}, we obtain
the modified Stefan--Boltzmann law taking into account the conformal anomaly as
\begin{align} \label{eq:modified_SB_law}
  \gamma T^2 = \varepsilon + \frac{1}{2} \langle T^\mu_\mu \rangle,
\end{align}
where $\gamma$ is an integration constant.
In the traceless limit, Eq.~\eqref{eq:modified_SB_law} reduces to the conventional two-dimensional Stefan--Boltzmann law~\cite{Tolman:1930zza}; however, the proper temperature would not
always be real, because the renormalized proper energy density may become negative near the horizon.
Finally, the proper temperature~\eqref{eq:modified_SB_law} can be written as
\begin{equation}
 \gamma T^2 = \frac{1}{f} \left( \langle T_{++} \rangle + \langle T_{--} \rangle \right) \label{eq:modified SB law 2}
\end{equation}
by using Eq.~\eqref{eq:epf2} and Eq.~\eqref{eq:trace_rel}.
For the two-dimensional $N$ massless scalar fields, the Stefan--Boltzmann constant is $\gamma=\frac{N\pi}{6}$~\cite{Christensen:1977jc}.
Thus, incorporating the conformal anomaly yields the modified Stefan–Boltzmann relation and the corresponding anomaly-induced proper temperature.

\section{The effective proper temperature for the quantum atmosphere}
\label{sec:temp}
The integration functions related to the thermal state can be determined by imposing the Hartle--Hawking state~\cite{Hartle:1976tp,Israel:1976ur},
which requires regularity of the stress tensor on the horizon.
In the conformal coordinates, the regularity condition is realized by imposing
$\langle T_{\pm\pm}\rangle\big|_{r=r_{\rm H}}=0$, which fixes the integration functions in Eq.~\eqref{eq:T++beta} as
\begin{equation} \label{eq:tHH}
    t^{\rm HH}_\pm = \frac{N}{192\pi}(f'(r_{\rm H}))^2 = \frac{N\pi}{12\beta_{\rm H}^2}.
\end{equation}
Then, the energy density in Eq.~\eqref{eq:epf2} is explicitly determined
as
\begin{equation} \label{eq:local energy}
\varepsilon_{\rm HH} = \frac{N}{96\pi f}\left(4ff''-f'^2 \right) + \frac{N\pi}{6f\beta_{\rm H}^2},
\end{equation}
which is not always positive~\cite{Visser:1996ix}.
Despite the negative energy density near the horizon in Eq.~\eqref{eq:local energy}, the proper temperature~\eqref{eq:modified_SB_law}
will be shown to be real in the next section since the trace term cancels the negative energy contribution.
Thus, the positive semidefinite proper temperature~\eqref{eq:modified SB law 2} in the Hartle--Hawking state is obtained as
\begin{equation} \label{eq:THH}
    \gamma T^2_{\rm HH}
    = \frac{1}{f} \left( \langle T_{++} \rangle_{\text{HH}} + \langle T_{--} \rangle_{\text{HH}} \right)
    = \frac{N}{96\pi f}\left( 2ff''-f'^2 +f'^2(r_\text{H}) \right),
\end{equation}
where $\langle T_{\pm\pm} \rangle_{\rm HH}$ are the stress tensors~\eqref{eq:T++beta} satisfying Eq.~\eqref{eq:tHH}.
The proper temperature~\eqref{eq:THH} vanishes at the horizon as a consequence of the cancellation between the state-dependent thermal contribution and the geometry-dependent vacuum-polarization contribution.
On the other hand, for the Boulware state~\cite{Boulware:1974dm} which is defined by the absence of radiation at infinity as $\langle T_{\pm\pm}\rangle |_{r\to\infty} = 0$, the integration functions are determined as $t^{\rm B}_\pm=0$. The corresponding energy density in Eq.~\eqref{eq:epf2} is
simply calculated as $\varepsilon_{\rm B} = \frac{N}{96\pi f}\left(4ff''-f'^2 \right)$.
It diverges negatively at the horizon and approaches zero from below at spatial infinity.
Note that the net flux is always zero in both the Hartle--Hawking state and the Boulware state.

An evaporating black hole can be described by imposing the Unruh state~\cite{Unruh:1976db}, constructed in such a way that the ingoing sector satisfies the Boulware boundary condition at infinity, whereas the outgoing sector satisfies the regularity condition appropriate to the Hartle–Hawking state at the future horizon.
To implement this boundary condition, we choose the integration functions as $t_+^{\rm U}=t_+^{\rm B}=0$ and $t_-^{\rm U}=t_-^{\rm HH}=\frac{N\pi}{12\beta^2_{\rm H}}$, yielding the stress tensors~\eqref{eq:T++beta} as $\langle T_{++}\rangle_{\rm U} = \langle T_{++}\rangle_{\rm B}$ and $\langle T_{--}\rangle_{\rm U} = \langle T_{--}\rangle_{\rm HH}$.
We now introduce the proper temperature $T_{\rm U}$ through the two-dimensional Stefan–Boltzmann relation in the radiating system~\cite{Giddings:2015uzr}
\begin{equation} \label{eq:SB_law_Unruh}
    \mathcal{F}
    =-\frac{1}{f(r)} \left(\langle T_{++} \rangle_{\text U} -\langle T_{--} \rangle_{\text U} \right)=\sigma \left( \frac{T_{\rm{H}}}{\sqrt{f(r)}} \right)^2
    = \sigma T^2_{\rm U}
\end{equation}
with $\sigma= \frac{\gamma}{2}$.
The squared proper temperature~\eqref{eq:SB_law_Unruh} can be decomposed into the left (ingoing) temperature and the right (outgoing) temperature as
\begin{equation} \label{eq:TLR}
    \sigma T_\text{L}^2 = -\frac{\langle T_{++}\rangle_{\text{U}}}{f(r)} = -\frac{\langle T_{++}\rangle_{\text{B}}}{f(r)},\qquad
    \sigma T_\text{R}^2 = \frac{\langle T_{--} \rangle_{\text{U}}}{f(r)} = \frac{\langle T_{--} \rangle_{\text{HH}}}{f(r)}.
\end{equation}
Here, the left and right fluxes are decoupled at the semiclassical level, allowing the ingoing and outgoing contributions to be characterized separately.
The left temperature in Eq.~\eqref{eq:TLR} diverges at the horizon and monotonically decreases to zero as $r\to\infty$.
However, the right temperature in Eq.~\eqref{eq:TLR} becomes finite everywhere.
Since $\langle T_{++}\rangle_{\text{HH}}=\langle T_{--}\rangle_{\text{HH}}$ in equilibrium, the right temperature in Eq.~\eqref{eq:TLR}
turns out to be equivalent to the proper temperature~\eqref{eq:THH}, namely
$T_{\rm R}= T_{\rm HH}$.
Henceforth, we identify the effective proper temperature for the quantum atmosphere with the right temperature characterizing the outgoing Hawking flux as
\begin{equation} \label{eq:eff}
T_{\rm eff}=T_{\rm R}= T_{\rm HH}.
\end{equation}
In this regard, the characteristics of the quantum atmosphere can be studied by calculating the maximum of $T_{\rm HH}$.

\begin{figure}[t]
\centering
\includegraphics[width=0.6\textwidth]{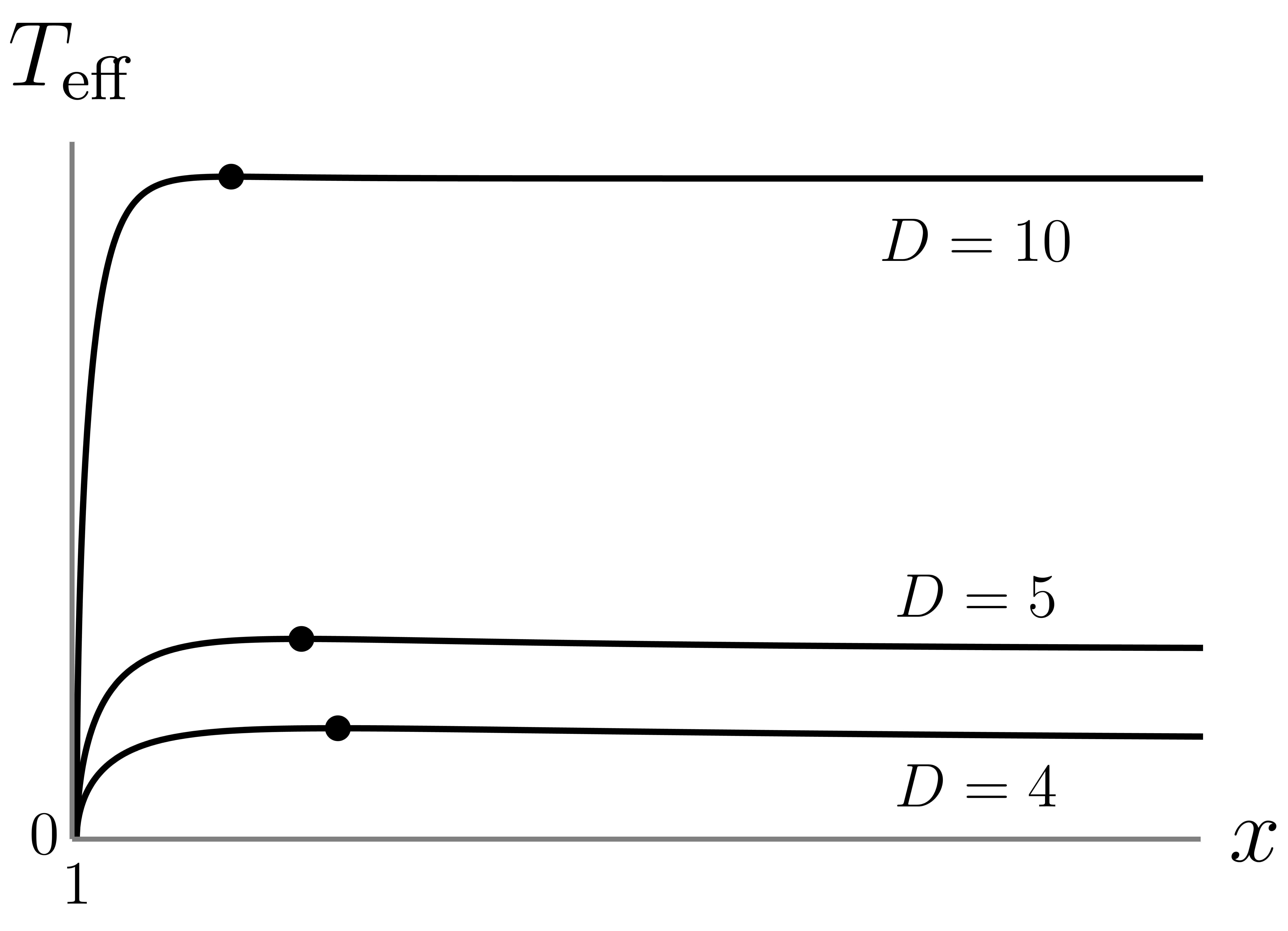}
\caption{The effective proper temperature~\eqref{eq:TR} is plotted for $D=4,5,10$, where $\mu=1$ for simplicity. The black dot indicates the location of the maximum of the effective proper temperature, where $x_{\rm atm}\approx 2.16, 2, 1.69$ for $D=4,5,10$, respectively. At $x=1$, $T_{\rm{eff}}=0$ for any $D$ and for $x\to\infty$
$T_{\rm{eff}} \to T_{\rm H}$, where $T_{\rm H} \approx 0.080, 0.16, 0.56$ for $D=4,5,10$, respectively.
}
\label{fig:Teff}
\end{figure}

\section{Quantum atmosphere in $D$-dimensional Schwarzschild black holes}
\label{sec:atmosphere}

\begin{figure}[t]
\centering
\includegraphics[width=0.6\textwidth]{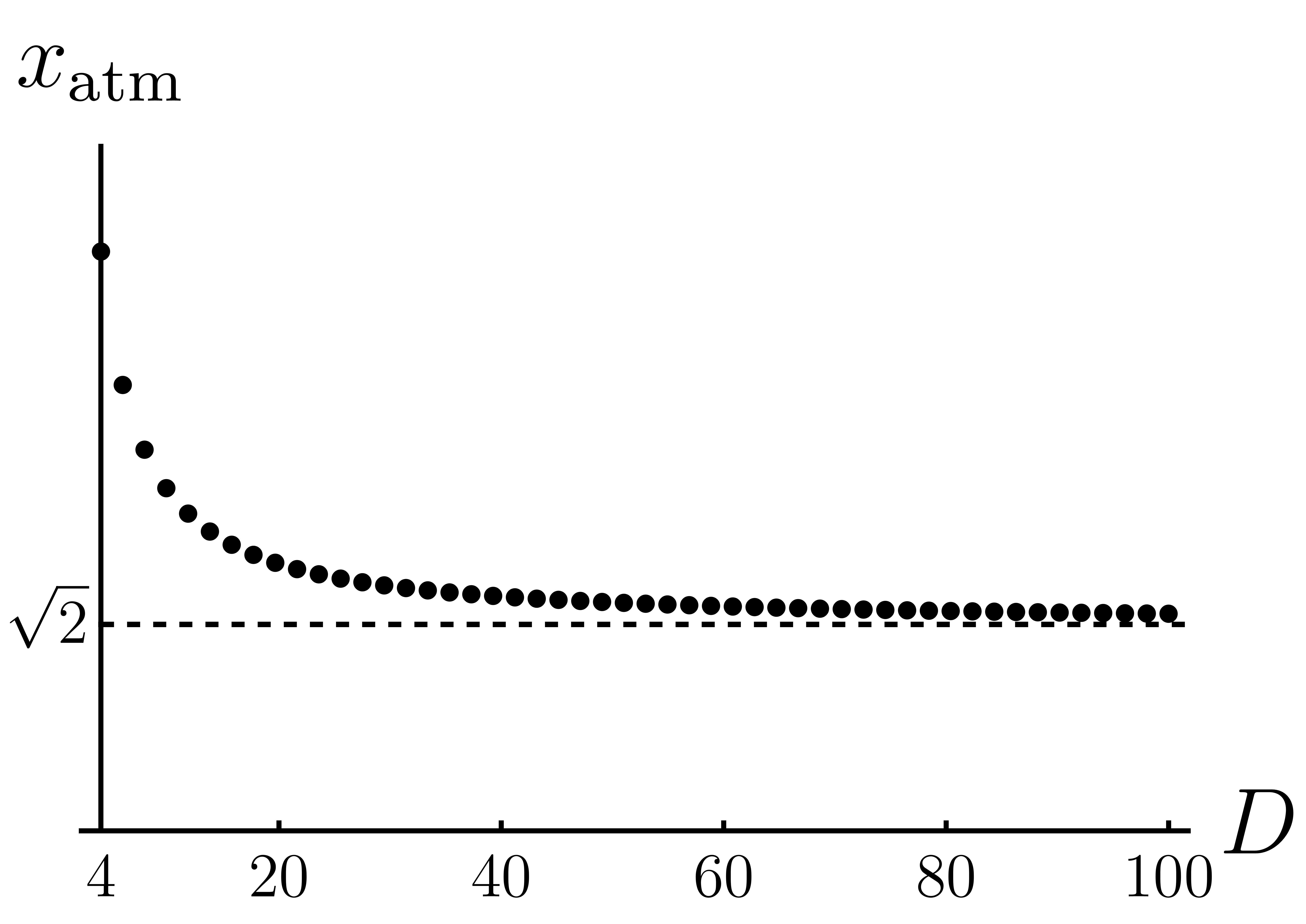}
\caption{For the quantum atmosphere, the dimensionless radius $x_{\rm atm}$ is plotted for $4\le D \le 100$. As $D$ increases,
the location of the maximum of $T_{\rm{eff}}$ decreases monotonically,
where $x_{\rm atm} \approx 2.16$ for $D=4$.
As $D\to \infty$, the value of $x_{\rm atm}$ approaches the constant ratio from the horizon, $x_{\rm atm}=\sqrt{2}$.}
\label{fig:xatm}
\end{figure}

We consider the two-dimensional radial sector of $D$-dimensional Schwarzschild black holes described by the metric function
\begin{eqnarray} \label{eq:metric}
f(r)= 1-\frac{\mu}{r^{D-3}}
\end{eqnarray}
where the mass parameter $\mu=\frac{16\pi G_{\rm D} M} {(D-2)\Omega_{D-2}}$, $\Omega_{D-2}=\frac{2\pi^{(D-1)/2}}{\Gamma\left((D-1)/2\right)}$,
and $D\ge4$.
The event horizon $r_{\rm H}$ is defined as $r_{\rm H}=\mu^{\frac{1}{D-3}}$ and
the Hawking temperature is given by the definition of the surface gravity as
$T_\text{H}=\frac{f'(r_\text{H})}{4\pi}=\frac{D-3}{4\pi r_{\rm H}}$~\cite{Hawking:1974sw}.
Plugging Eq.~\eqref{eq:metric} into Eq.~\eqref{eq:eff}, we obtain
the effective proper temperature for the quantum atmosphere,
\begin{align}
    T_{\rm eff}&=\frac{T_{\rm H}}{\sqrt{D-3}} \sqrt{\frac{x^{D-3}}{x^{D-3}-1} \left( D-3 - \frac{2(D-2)}{x^{D-1}} + \frac{D-1}{x^{2(D-2)}} \right)}, \label{eq:TR}
\end{align}
where the dimensionless radius is introduced as $x= \frac{r}{r_{\rm H}}$
with $x\ge1$.
The behavior of the effective proper temperature~\eqref{eq:TR} is illustrated in Fig.~\ref{fig:Teff} for various dimensions.
For any $D$, the effective proper temperature vanishes at the horizon where $x=1$.
As $x$ increases, the effective proper temperature rises to its maximum at $x=x_{\rm atm}$ and then decreases towards the corresponding $D$-dimensional Hawking temperature as $x\to\infty$.
In addition, for $x>x_{\rm atm}$, $T_{\rm eff} \to T_H$ as $D$ increases, indicating that the contrast between the maximum and the asymptotic value becomes increasingly small.

Let us define the atmospheric radius for the quantum atmosphere
as the radial position $x=x_{\rm atm}$ at which the effective proper temperature~\eqref{eq:TR} attains its maximum.
This position is determined by solving
\begin{equation} \label{eq:atm}
    (D-3)^2 x_{\rm atm}^{2(D-2)} - 2(D-2)(D-1)x_{\rm atm}^{2(D-3)} + 2(D-2)(D+1)x_{\rm atm}^{D-3} - (D-1)^2 = 0.
\end{equation}
For $D=4$ and $D=5$, Eq.~\eqref{eq:atm} admits simple analytic solutions such as  $x_{\rm atm}=\sqrt{10}-1\approx 2.16$ and $x_{\rm atm}=2$, respectively.
For higher dimensions, however, Eq.~\eqref{eq:atm} must in general be solved numerically.
In Fig.~\ref{fig:xatm}, the numerical values of $x_{\rm atm}$ for various dimensions are plotted.
As $D$ increases, $x_{\rm atm}$ decreases in a manner similar to that shown in Ref.~\cite{Hod:2016hdd}; however, $x_{\rm atm}$ approaches the asymptotic value.
In the large-dimensional limit,
it becomes
\begin{equation} \label{eq:atm_largeD}
    x_{\rm atm} \sim \sqrt{2} + \mathcal{O}\left(\frac{1}{D}\right) \to \sqrt{2},
\end{equation}
where the atmospheric radius remains separated from the horizon radius by a factor of $\sqrt{2} r_{\rm H}$.
Consequently, the atmospheric radius remains separated from the horizon by a finite radial factor, indicating that the quantum atmosphere can persist as an extended exterior region in any dimension.

\section{Conclusion and discussion}
\label{sec:conclusion}

In this paper, we investigated the
effective proper temperature of the black-hole quantum atmosphere in the two-dimensional radial sector of $D$-dimensional Schwarzschild black holes.
Starting from the renormalized stress tensor for $N$ massless scalar fields on a general static two-dimensional background,
we derived the modified Stefan--Boltzmann law in the presence of the conformal anomaly and the corresponding proper temperature in the Hartle--Hawking state.
From the flux in the Unruh state, we identified the effective proper temperature for the outgoing flux with the proper temperature in the Hartle--Hawking state.
Using the effective proper temperature for the two-dimensional radial sector of $D$-dimensional Schwarzschild black holes, we defined the atmospheric radius as the location of the maximum of the effective proper temperature.
We showed that the atmospheric radius decreases as the spacetime dimension increases but remains separated from the horizon in the large-dimensional limit.
Our result suggests that, from the local stress-tensor perspective, the spatial origin of the outgoing Hawking radiation remains an extended exterior region even in the large-dimensional limit.

One may ask how the conventional Tolman temperature is related to the proper temperatures in the present work.
For the traceless stress tensor, as discussed in Ref.~\cite{Tolman:1930zza}, Eq.~\eqref{eq:modified_SB_law}
reduces to the conventional Stefan-Boltzmann law $\gamma T^2=\varepsilon$.
In the Hartle--Hawking state, the proper energy density
$\varepsilon$ is quantum-mechanically negative near the horizon
due to the negative contribution
of vacuum polarization of $\hat{\varepsilon}$.
Thus, if the geometry-dependent contribution from $\hat{\varepsilon}$
can be neglected, then $\gamma T^2 = t(r)$.
Substituting the boundary condition \eqref{eq:tHH} into Eq.~\eqref{eq:t_def},
one can reproduce the familiar Tolman form $T=\frac{T_{\rm H}}{\sqrt{f}}$.
Now, it is worth noting that the decomposed stress tensors of $\hat{T}_{\pm\pm}$ and $t_\pm$ are coordinate dependent: neither contribution should be interpreted as an independently covariant stress tensor, although their sum transforms as the true tensor. In particular, the state-dependent functions $t_\pm$ may be identified with the expectation values of the corresponding normal-ordered stress tensors because they are linked to the particular coordinate system. Thus,
$t(r)$ cannot be a true tensor~\cite{Fabbri:2005mw}.
In contrast, Eq.~\eqref{eq:modified_SB_law} is expressed in terms of the proper energy density and the scalar trace and therefore defines a coordinate-invariant temperature for a specified observer.
On the other hand, in the Unruh state, Eq.~\eqref{eq:SB_law_Unruh} tells us that the Tolman form of the temperature is manifestly coordinate invariant since the
combination
of integration functions, $t_+ - t_-$, is shown to be a covariant tensor because of the cancellation of anomalous terms under coordinate transformations.
Thus, the Tolman form is
not proper in the Hartle--Hawking state whereas it is proper in the Unruh state.

\acknowledgments
We have benefited from discussions with M. Eune.
This research was supported by Basic Science Research Program
through the National Research Foundation of Korea (NRF) funded by the Ministry
of Education through the Center for Quantum Spacetime (CQUeST) of Sogang
University (No. RS-2020-NR049598).  This work was supported by the National
Research Foundation of Korea (NRF) grant funded by the Korea government (MSIT)
(No. RS-2022-NR069013).

\bibliographystyle{JHEP}       
\bibliography{references}

\end{document}